\documentclass[letterpaper, 10 pt, conference]{ieeeconf}  

\IEEEoverridecommandlockouts                              

\usepackage{graphics} 
\usepackage{epsfig} 
\usepackage{mathptmx} 
\usepackage{times} 
\usepackage{amsmath} 
\usepackage{amssymb}  
\usepackage[export]{adjustbox}

\title{\LARGE \bf
Learning to Decode Linear Codes Using Deep Learning
}

\author{Eliya~Nachmani$^{1}$, Yair~Be'ery$^{2}$ and David~Burshtein$^{3}$ %
\thanks{$^{1}$Eliya Nachmani is with the School of Electrical Engineering, Tel-Aviv University,
{\tt\small enk100@gmail.com}}%
\thanks{$^{2}$Yair~Be'ery is with the School of Electrical Engineering, Tel-Aviv University,
{\tt\small ybeery@eng.tau.ac.il}}%
\thanks{$^{3}$David~Burshtein is with the School of Electrical Engineering, Tel-Aviv University,
{\tt\small burstyn@eng.tau.ac.il}}%
}

\begin{document}

\maketitle
\thispagestyle{empty}
\pagestyle{empty}

\begin{abstract}

A novel deep learning method for improving the belief propagation algorithm is proposed. The method generalizes the standard belief propagation algorithm by assigning weights to the edges of the Tanner graph. These edges are then trained using deep learning techniques. A well-known property of the belief propagation algorithm is the independence of the performance on the transmitted codeword. A crucial property of our new method is that our decoder preserved this property. Furthermore, this property allows us to learn only a single codeword instead of exponential number of codewords. Improvements over the belief propagation algorithm are demonstrated for various high density parity check codes.

\end{abstract}

\section{INTRODUCTION}
In recent years deep learning methods have demonstrated significant improvements in various tasks. These methods outperform human-level object detection in some tasks \cite{resnet}, and achieve state-of-the-art results in machine translation \cite{nmt} and speech processing \cite{graves2013speech}. Additionally, deep learning combined with reinforcement learning techniques was able to beat human champions in challenging games such as Go \cite{d_silver}. There are three different reasons for the outstanding results of Deep Learning models:
\begin{enumerate}
\item Powerful computing resources such as fast GPUs.
\item Utilizing efficiently large collections of datasets, e.g. ImageNet \cite{imagenet} for image processing.
\item Advanced academic research on training methods and network architectures \cite{batch_norm}, \cite{alexnet}, \cite{adadelta}, \cite{dropout}.
\end{enumerate}
Error correcting codes for channel coding are used in order to enable reliable communications at rates close to the Shannon capacity. A well-known family of linear error correcting codes are the low-density parity-check (LDPC) codes \cite{galmono}. LDPC codes achieve near Shannon channel capacity with the belief propagation (BP) decoding algorithm, but can typically do so for relatively large block lengths. For high density parity check (HDPC) codes \cite{jiang2006iterative}, \cite{dimnik2009improved}, \cite{yufit2011efficient}, \cite{zhang2012adaptive}, such as common powerful algebraic codes, the BP algorithm obtains poor results compared to the maximum likelihood decoder \cite{helmling2014efficient}.
In this work we focus on HDPC codes and demonstrate how the BP algorithm can be improved. The naive approach to the problem is to assume a neural network type decoder without restrictions, and train its weights using a dataset that contains a large amount of codewords. The training goal is to reconstruct the transmitted codeword from a noisy version after transmitting over the communication channel. Unfortunately, when using this approach our decoder is not given any side information regarding the structure of the code. In fact it is even not aware of the fact that the code is linear. Hence we are required to train the decoder using a huge collection of codewords from the code, and due to the exponential nature of the problem, this is infeasible, e.g., for a BCH(63,45) code we need a dataset of $2^{45}$ codewords. On top of that, the database needs to reflect the variability due to the noisy channel.
In order to overcome this issue, our proposed approach is to assign weights to the edges of the Tanner graph that represent the given linear code, thus yielding a ``soft'' Tanner graph. These edges are trained using deep learning techniques. A well-known property of the BP algorithm is the independence of the performance on the transmitted codeword. A major ingredient in our new method is that this property is preserved by our decoder. Thus it is sufficient to use a single codeword for training the parameters of our decoder. We demonstrate improvements over BP for various high density parity check codes, including BCH(63,36), BCH(63,45), and BCH(127,106) .

\section{THE BELIEF PROPAGATION ALGORITHM}

The renowned BP decoder \cite{galmono}, \cite{ru_book} can be constructed from the Tanner graph, which is a graphical representation of some parity check matrix that describes the code. In this algorithm, messages are transmitted over edges. Each edge calculates its outgoing message based on all incoming messages it receives over all its edges, except for the message received on the transmitting edge. We start by providing an alternative graphical representation to the BP algorithm with $L$ full iterations when using parallel (flooding) scheduling. Our alternative representation is a trellis in which the nodes in the hidden layers correspond to edges in the Tanner graph.  
Denote by $N$, the code block length (i.e., the number of variable nodes in the Tanner graph), and by $E$, the number of edges in the Tanner graph. Then the input layer of our trellis representation of the BP decoder is a vector of size $N$, that consists of the log-likelihood ratios (LLRs) of the channel outputs. The LLR value of variable node $v$, $v=1,2,\ldots,N$, is given by
$$
l_v = \log\frac{\Pr\left(C_v=1 | y_v\right)}{\Pr\left(C_v=0 | y_v\right)}
$$
where $y_v$ is the channel output corresponding to the $v$th codebit, $C_v$. 

All the following layers in the trellis, except for the last one (i.e., all the hidden layers), have size $E$. For each hidden layer, each processing element in that layer is associated with the message transmitted over some edge in the Tanner graph. The last (output) layer of the trellis consists of $N$ processing elements that output the final decoded codeword. Consider the $i$th hidden layer, $i=1,2,\ldots,2L$. For odd (even, respectively) values of $i$, each processing element in this layer outputs the message transmitted by the BP decoder over the corresponding edge in the graph, from the associated variable (check) node to the associated check (variable) node. A processing element in the first hidden layer ($i=1$), corresponding to the edge $e=(v,c)$, is connected to a single input node in the input layer: It is the variable node, $v$, associated with that edge.
Now consider the $i$th ($i>1$) hidden layer. For odd (even, respectively) values of $i$, the processing node corresponding to the edge $e=(v,c)$ is connected to all processing elements in layer $i-1$ associated with the edges $e'=(v,c')$ for $c'\ne c$ ($e'=(v',c)$ for $v'\ne v$, respectively). For odd $i$, a processing node in layer $i$, corresponding to the edge $e=(v,c)$, is also connected to the $v$th input node.

The BP messages transmitted over the trellis graph are the following. Consider hidden layer $i$, $i=1,2,\ldots,2L$, and let $e=(v,c)$ be the index of some processing element in that layer. We denote by $x_{i,e}$, the output message of this processing element. For odd (even, respectively), $i$, this is the message produced by the BP algorithm after $\lfloor (i-1)/2 \rfloor$ iterations, from variable to check (check to variable) node.

For odd $i$ and $e=(v,c)$ we have (recall that the self LLR message of $v$ is $l_v$),
\begin{equation}
x_{i,e=(v,c)} = l_v + \sum_{e'=(v,c'),\: c'\ne c} x_{i-1,e'}
\label{eq:x_ie_RB}
\end{equation}
under the initialization, $x_{0,e'}=0$ for all edges $e'$ (in the beginning there is no information at the parity check nodes). The summation in~\eqref{eq:x_ie_RB} is over all edges $e'=(v,c')$ with variable node $v$ except for the target edge $e=(v,c)$. Recall that this is a fundamental property of message passing algorithms~\cite{ru_book}.

Similarly, for even $i$ and $e=(v,c)$ we have,
\begin{equation}
x_{i,e=(v,c)} = 2\tanh^{-1} \left( \prod_{e'=(v',c),\: v'\ne v} \tanh \left( \frac{x_{i-1,e'}}{2} \right) \right)
\label{eq:x_ie_LB}
\end{equation}

The final $v$th output of the network is given by
\begin{equation}
o_v = l_v + \sum_{e'=(v,c')} x_{2L,e'}
\label{eq:ov}
\end{equation}
which is the final marginalization of the BP algorithm.

\section{THE PROPOSED DEEP NEURAL NETWORK DECODER}

We suggest the following parameterized deep neural network decoder that generalizes the BP decoder of the previous section. We use the same trellis representation for the decoder as in the previous section. The difference is that now we assign weights to the edges in the Tanner graph. These weights will be trained using stochastic gradient descent which is the standard method for training neural networks. More precisely, our decoder has the same trellis architecture as the one defined in the previous section. However, Equations~\eqref{eq:x_ie_RB}, \eqref{eq:x_ie_LB} and \eqref{eq:ov} are replaced by
\begin{equation}
x_{i,e=(v,c)} = \tanh \left(\frac{1}{2}\left(w_{i,v} l_v + \sum_{e'=(v,c'),\: c'\ne c} w_{i,e,e'} x_{i-1,e'}\right)\right)
\label{eq:x_ie_RB_NN}
\end{equation}
for odd $i$,
\begin{equation}
x_{i,e=(v,c)} = 2\tanh^{-1} \left( \prod_{e'=(v',c),\: v'\ne v}{x_{i-1,e'}}\right)
\label{eq:x_ie_LB_NN}
\end{equation}
for even $i$, and
\begin{equation}
o_v = \sigma \left( w_{2L+1,v} l_v + \sum_{e'=(v,c')} w_{2L+1,v,e'} x_{2L,e'} \right)
\label{eq:ov_NN}
\end{equation}
where $\sigma(x) \equiv \left( 1+e^{-x} \right)^{-1}$ is a sigmoid function. The sigmoid is added so that the final network output is in the range $[0,1]$. This makes it possible to train the network using a cross entropy loss function, as described in the next section. Apart of the addition of the sigmoid function at the outputs of the network, one can see that by setting all weights to one, Equations \eqref{eq:x_ie_RB_NN}-\eqref{eq:ov_NN} degenerate to \eqref{eq:x_ie_RB}-\eqref{eq:ov}. Hence by optimal setting (training) of the parameters of the neural network, its performance can not be inferior to plain BP.

It is easy to verify that the proposed message passing decoding algorithm \eqref{eq:x_ie_RB_NN}-\eqref{eq:ov_NN} satisfies the message passing symmetry conditions \cite[Definition 4.81]{ru_book}. Hence, by \cite[Lemma 4.90]{ru_book}, when transmitting over a binary memoryless symmetric (BMS) channel, the error rate is independent of the transmitted codeword. Therefore, to train the network, it is sufficient to use a database which is constructed by using noisy versions of a single codeword. For convenience we use the zero codeword, which must belong to any linear code. The database reflects various channel output realizations when the zero codeword has been transmitted. The goal is to train the parameters $\left \{ w_{i,v},w_{i,e,e'},w_{i,v,e'} \right \}$ to achieve an $N$ dimensional output word which is as close as possible to the zero codeword. The network architecture is a non-fully connected neural network. We use stochastic gradient descent to train the parameters.
The motivation behind the new proposed parameterized decoder is that by setting the weights properly, one can compensate for small cycles in the Tanner graph that represents the code. That is, messages sent by parity check nodes to variable nodes can be weighted, such that if a message is less reliable since it is produced by a parity check node with a large number of small cycles in its local neighborhood, then this message will be attenuated properly.

\begin{figure}[thpb]
	\centering
    \includegraphics[width=0.983101925\linewidth]{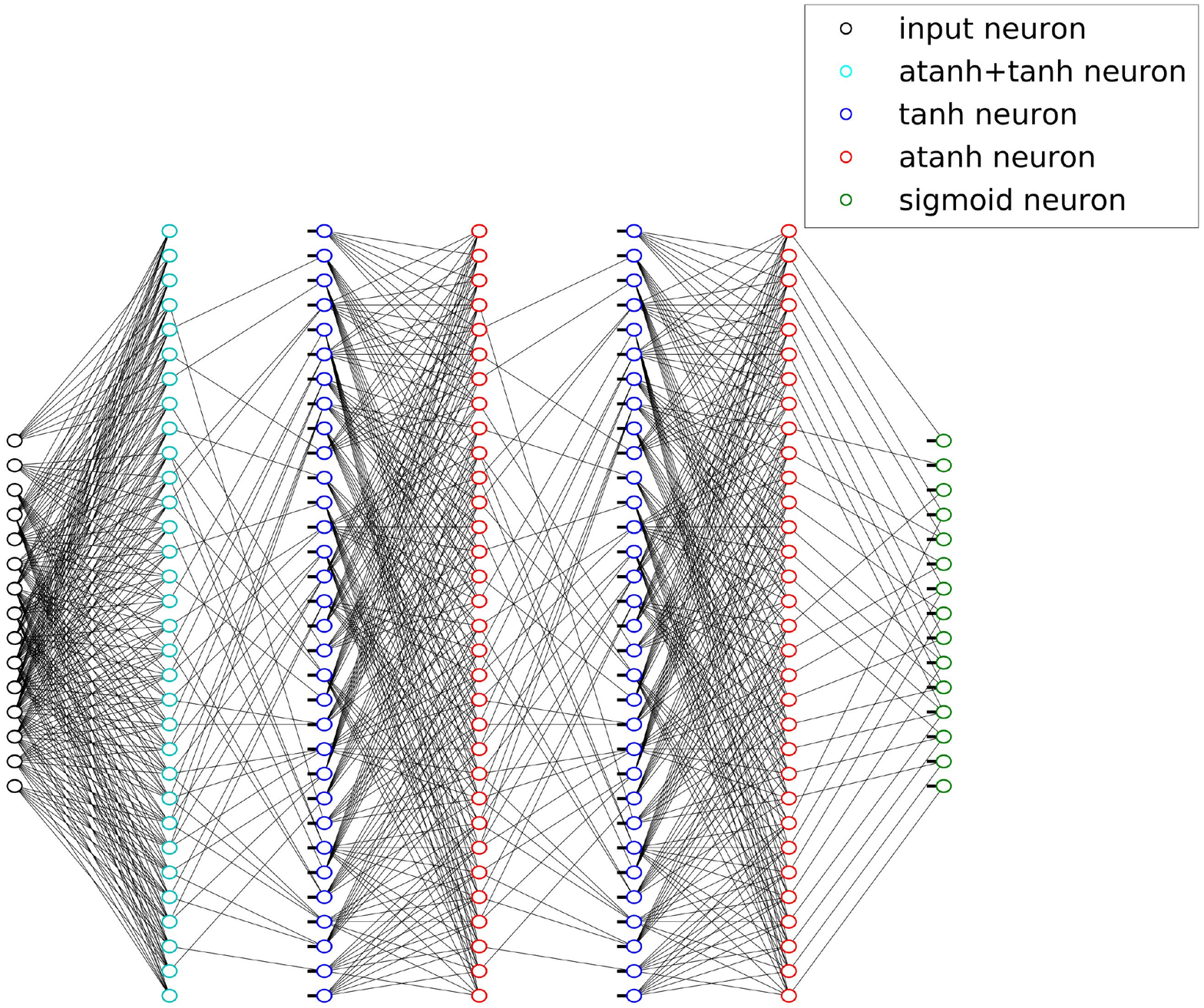}
	\caption{Deep Neural Network Architecture For BCH(15,11) with 5 hidden layers which correspond to 3 full BP iterations. Note that the self LLR messages $l_v$ are plotted as small bold lines. The first hidden layer and the second hidden layer that were described above are merged together. Also note that this figure shows 3 full iterations and the final marginalization.}
	\label{fig:BCH_15_11_arch}
\end{figure}

The time complexity of the deep neural network is similar to plain BP algorithm. Both have the same number of layers and the same number of non-zero weights in the Tanner graph. The deep neural network architecture is illustrated in Figure~\ref{fig:BCH_15_11_arch} for a BCH(15,11) code.

\section{EXPERIMENTS}

\subsection{Neural Network Training}
We built our neural network on top of the TensorFlow framework~\cite{abadi2015tensorflow} and used an NVIDIA Tesla K40c GPU for accelerated training. We applied cross entropy as our loss function, 
\begin{equation}
L{(o,y)}=-\frac{1}{N}\sum_{v=1}^{N}y_{v}\log(o_{v})+(1-y_{v})\log(1-o_{v})
\label{eq:cross_entropy}
\end{equation} 
where $o_{v}$, $y_{v}$ are the deep neural network output and the actual $v$th component of the transmitted codeword (if the all-zero codeword is transmitted then $y_{v}=0$ for all $v$).
Training was conducted using stochastic gradient descent with mini-batches. The mini-batch size was $120$ examples. We applied the RMSPROP~\cite{rmsprop} rule with a learning rate equal to $0.001$. The neural network has $10$ hidden layers, which correspond to $5$ full iterations of the BP algorithm. Each processing element in an odd (even, respectively) indexed hidden layer is described by Equation~\eqref{eq:x_ie_RB_NN} (Equation~\eqref{eq:x_ie_LB_NN}, respectively).
At test time, we inject noisy codewords after transmitting through an AWGN channel and measure the bit error rate (BER) in the decoded codeword at the network output.
When computing~\eqref{eq:x_ie_RB_NN}, we also clip the input to the tanh function such that the absolute value of the input is always smaller than some positive constant $A < 10$. This is also required for practical (finite block length) implementations of the BP algorithm, in order to stabilize the operation of the decoder. We trained our decoding network on few different linear codes, including BCH(15,11), BCH(63,36), BCH(63,45) and BCH(127,106).

\subsection{Neural Network Training With Multiloss}
The proposed neural network architecture has the property that after every odd $i$ layer we can add final marginalization. We can use that property to add additional terms in the loss. The additional terms can increase the gradient update at the backpropagation algorithm and allow learning the lower layers. At each odd $i$ layer we add the final marginalization to loss:
\begin{equation}
L{(o,y)}=-\frac{1}{N}\sum_{i=1,3}^{2L-1}\sum_{v=1}^{N}y_{v}\log(o_{v,i})+(1-y_{v})\log(1-o_{v,i})
\label{eq:multiloss_cross_entropy}
\end{equation}
where $o_{v,i}$, $y_{v}$ are the deep neural network output at the odd $i$ layer and the actual $v$th component of the transmitted codeword. This network architecture is illustrated in Figure 2.
\begin{figure}[thpb]
	\centering
    \hspace*{-3.5cm}\includegraphics[width=1.4\linewidth, left]{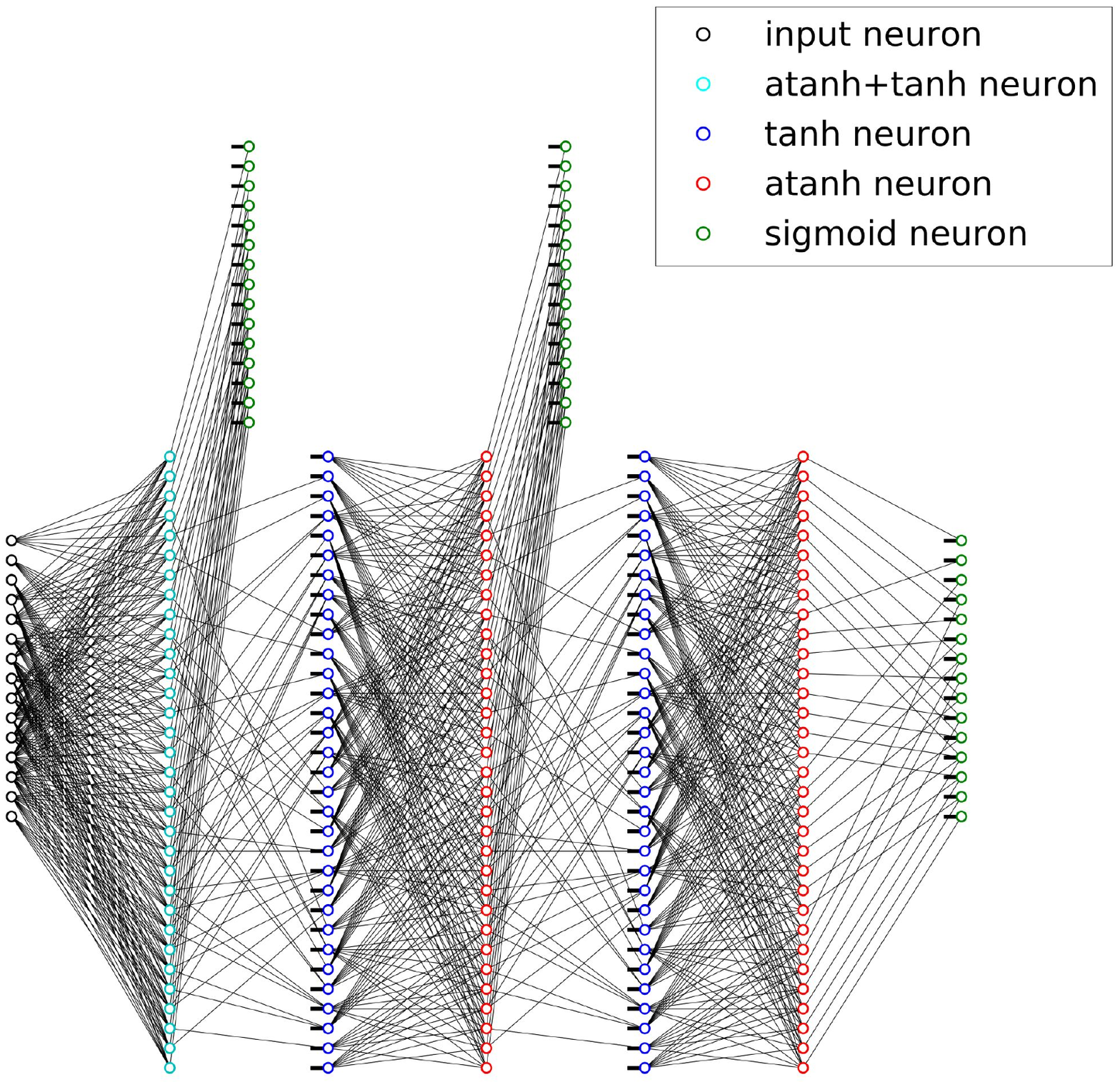}
    \caption{Deep Neural Network Architecture For BCH(15,11) with training
    multiloss. Note that the self LLR messages $l_v$ are plotted as small bold lines.}
	\label{fig:BCH_15_11_arch_multiloss}
\end{figure}

\subsection{Dataset}
The training data is created by transmitting the zero codeword through an AWGN channel with varying SNRs ranging from $1{\rm dB}$ to $6{\rm dB}$. Each mini batch has $20$ codewords for each SNR (a total of $120$ examples in the mini batch). For the test data we use codewords with the same SNR range as in the training dataset. Parity check matrices were taken from \cite{ParityCheckMatrix}.

\subsection{Results}
\begin{figure}[thpb]
	\centering
	\includegraphics[width=0.983101925\linewidth]{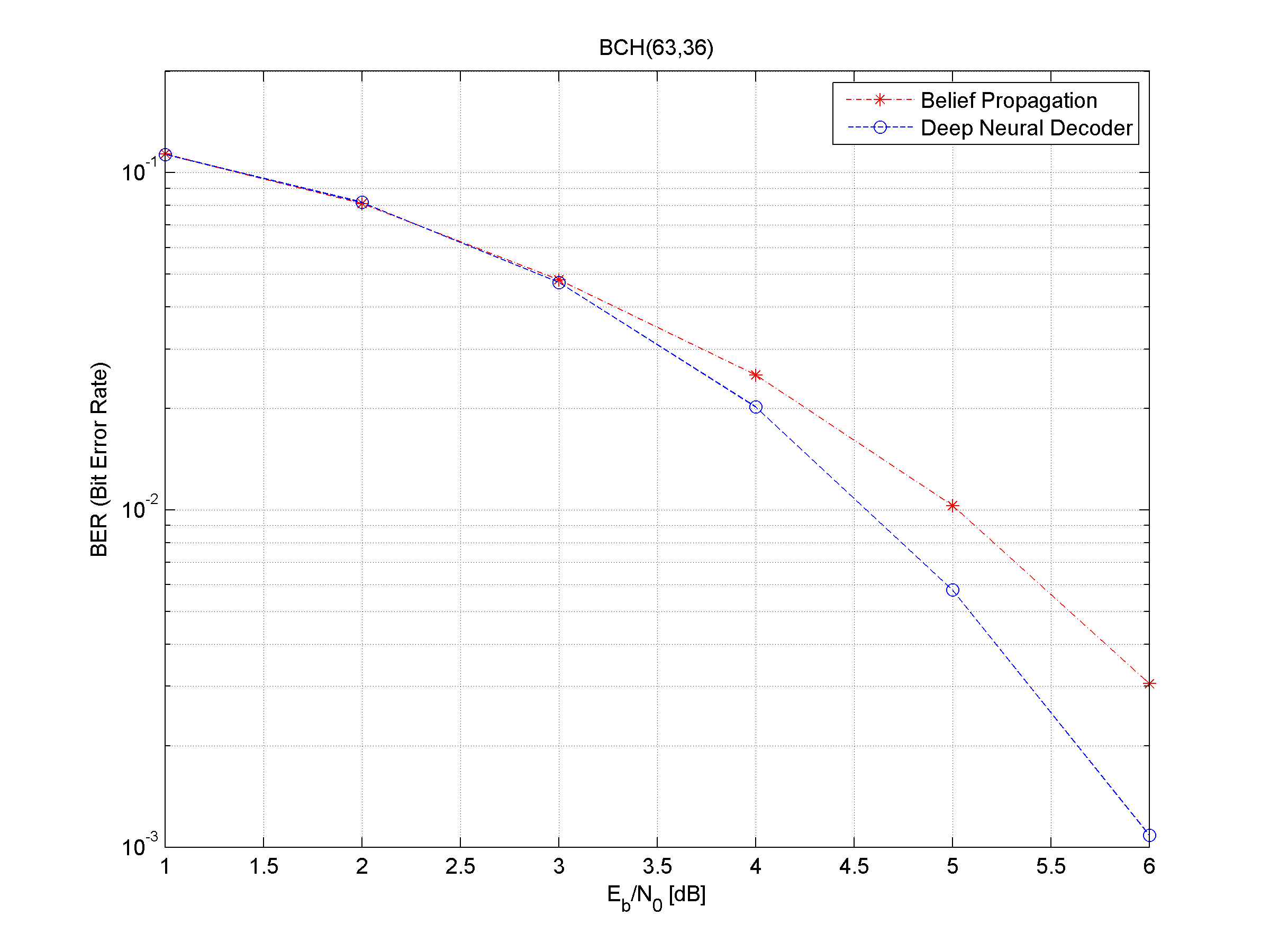}
	\caption{BER results for BCH(63,36) code.}
	\label{fig:BCH_63_36_ber}
\end{figure}

\begin{figure}[thpb]
	\centering
	\includegraphics[width=0.983101925\linewidth]{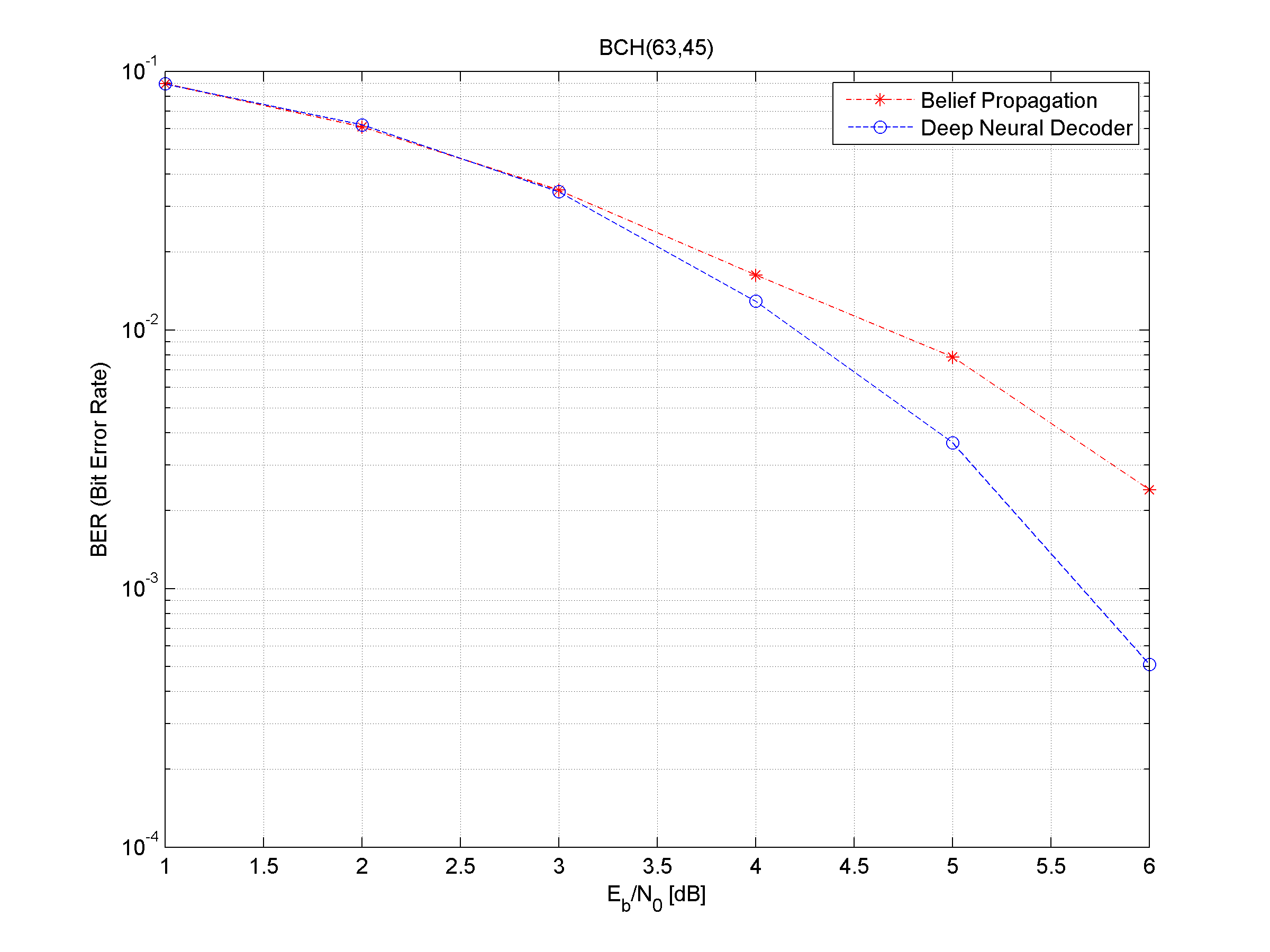}
	\caption{BER results for BCH(63,45) code.}
	\label{fig:BCH_63_45_ber}
\end{figure}

\begin{figure}[thpb]
	\centering
	\includegraphics[width=0.983101925\linewidth]{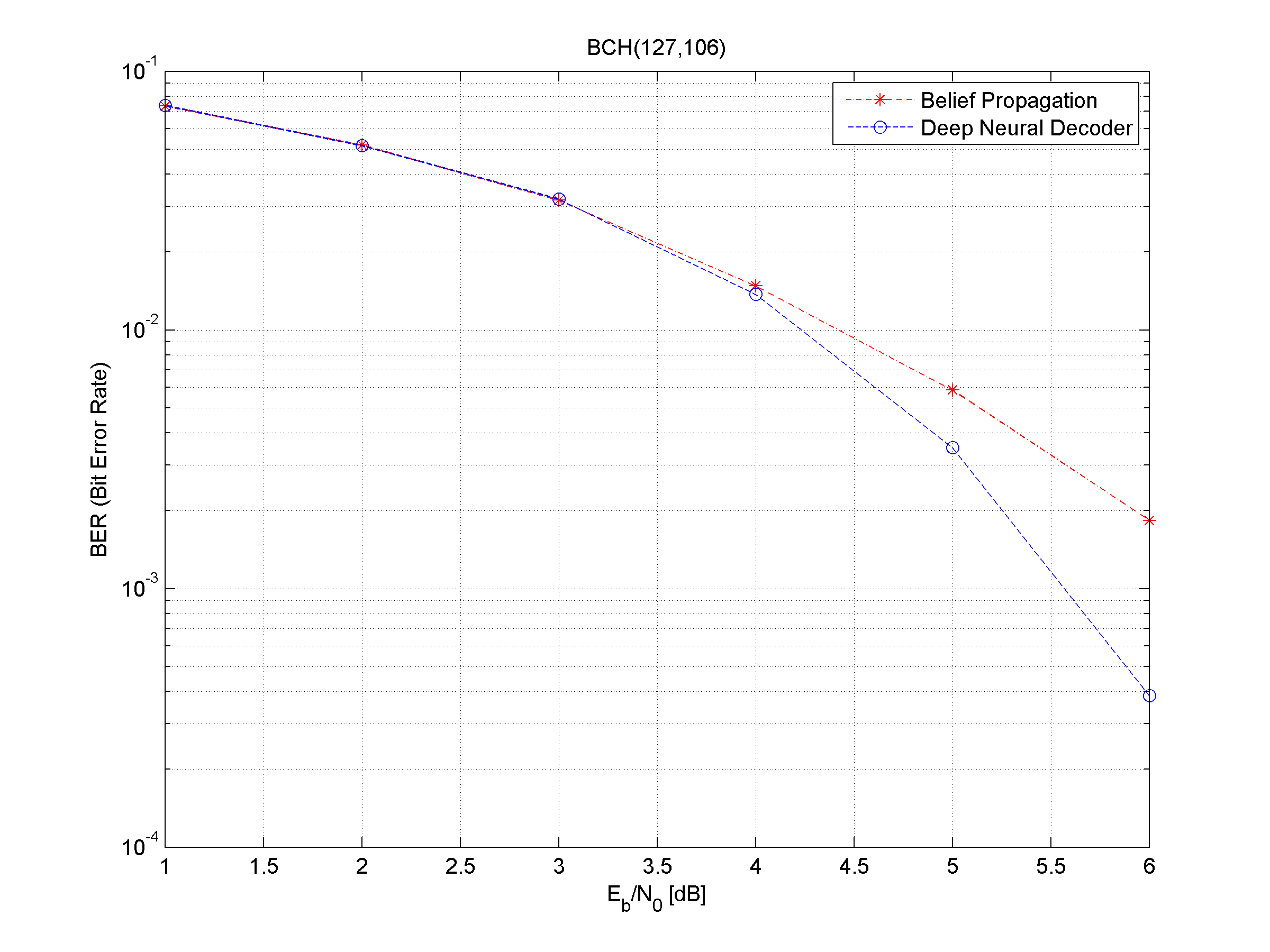}
	\caption{BER results for BCH(127,106) code.}
	\label{fig:BCH_127_106_ber}
\end{figure}

\begin{figure}[thpb]
	\centering
	\includegraphics[width=0.983101925\linewidth]{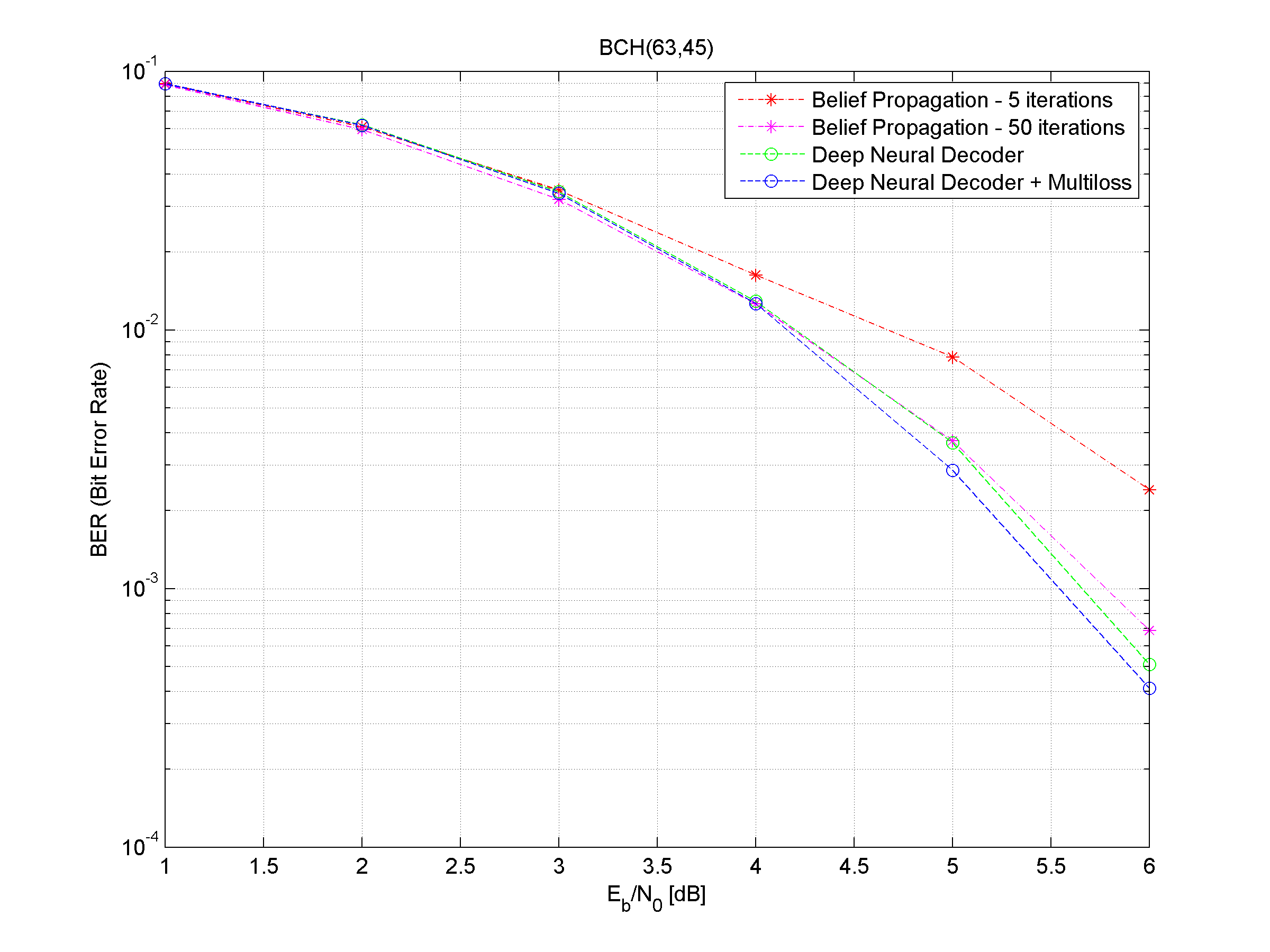}
	\caption{BER results for BCH(63,45) code trained with multiloss.}
	\label{fig:BCH_63_45_multiloss_ber}
\end{figure}

In this section we present the results of the deep neural decoding networks for various BCH block codes. In each code we observed an improvement compared to the BP algorithm. Note that when we applied our algorithm to the BCH(15,11) code, we obtained close to maximum likelihood results with the deep neural network. For larger BCH codes, the BP algorithm and the deep neural network still have a significant gap from maximum likelihood. The BER figures~\ref{fig:BCH_63_36_ber}, \ref{fig:BCH_63_45_ber} and \ref{fig:BCH_127_106_ber} show an improvement of up to $0.75{\rm dB}$ in the high SNR region. Furthermore, the deep neural network BER is consistently smaller or equal to the BER of the BP algorithm. This result is in agreement with the observation that our network cannot perform worse than the BP algorithm. 
Figure \ref{fig:BCH_63_45_multiloss_ber} shows the results of training the deep neural network with multiloss. It shows an improvement of up to $0.9{\rm dB}$ compared to the plain BP algorithm. Moreover, we can observe that we can achieve the same BER performance of 50 iteration BP with 5 iteration of the deep neural decoder, This is equal to complexity reduction of factor 10. 

\begin{figure}[thpb]
	\centering
	\includegraphics[width=0.983101925\linewidth]{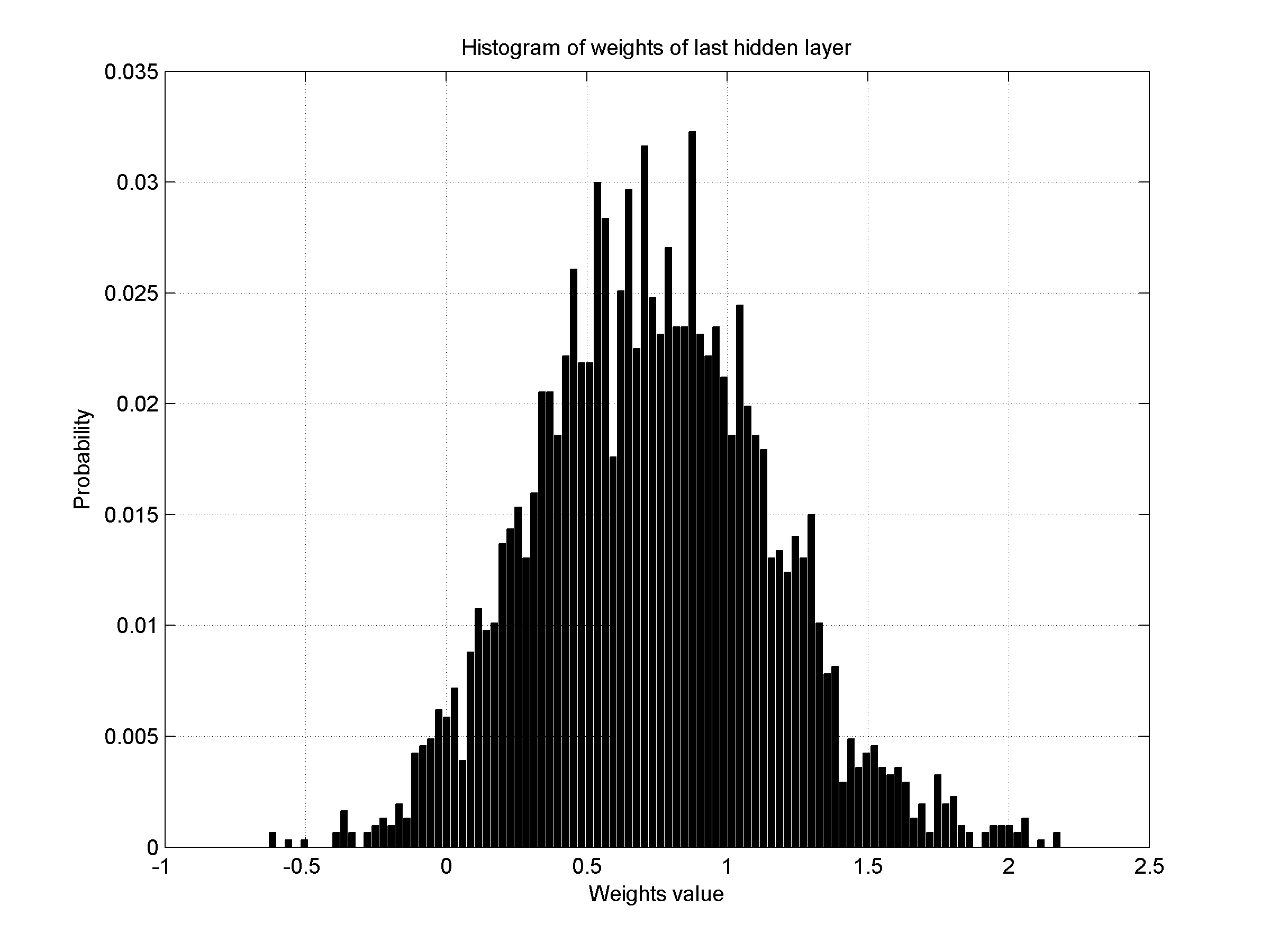}
	\caption{Weights histogram of last hidden layer of the deep neural network for BCH(63,45) code.}
	\label{fig:weight_hist}
\end{figure}

\begin{figure}[thpb]
	\centering
	\includegraphics[width=0.983101925\linewidth]{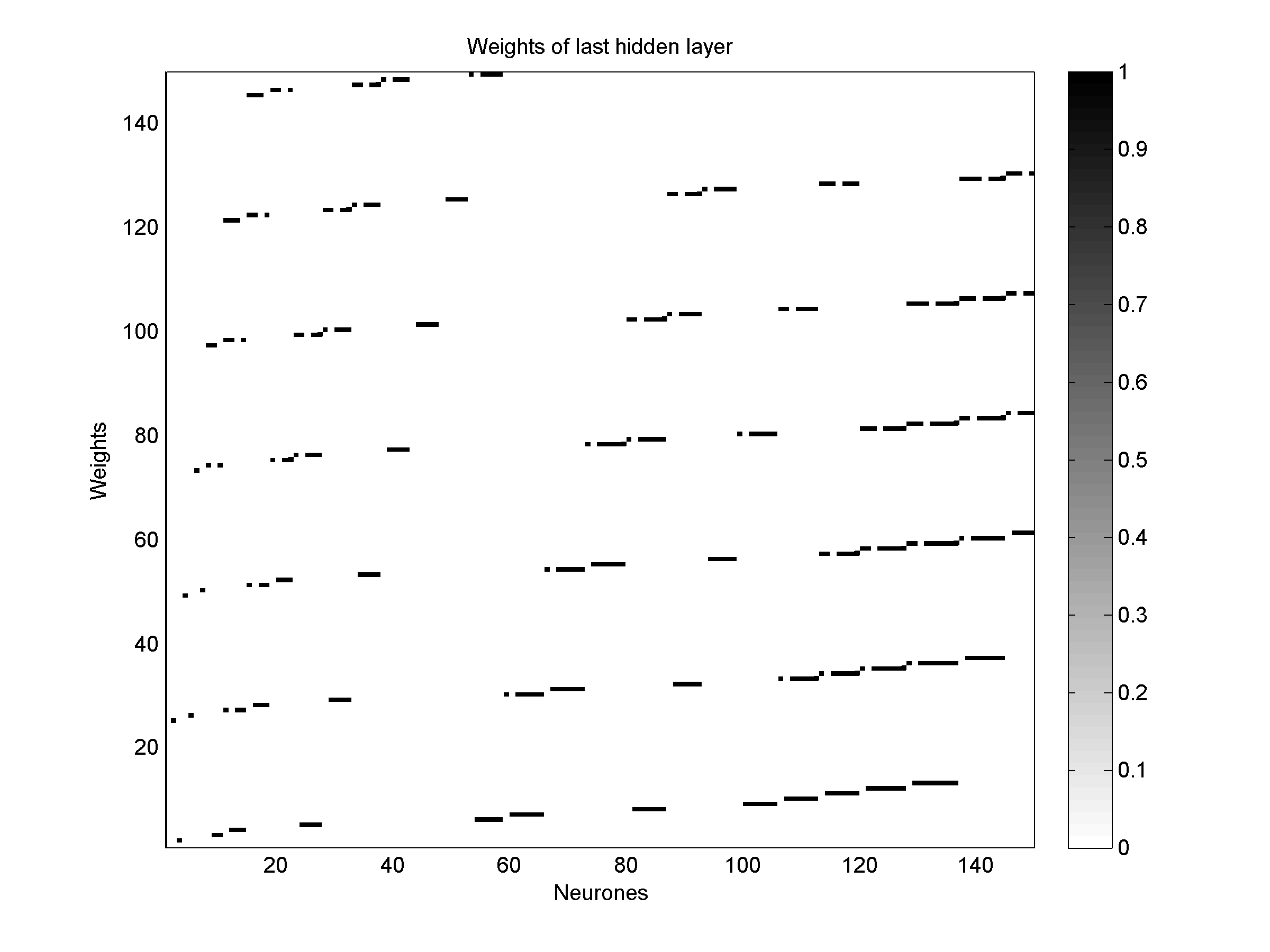}
	\caption{Weights of the last hidden layer in the BP algorithm for BCH(63,45) code.}
	\label{fig:layer4_bp}
\end{figure}

\begin{figure}[thpb]
	\centering
	\includegraphics[width=0.983101925\linewidth]{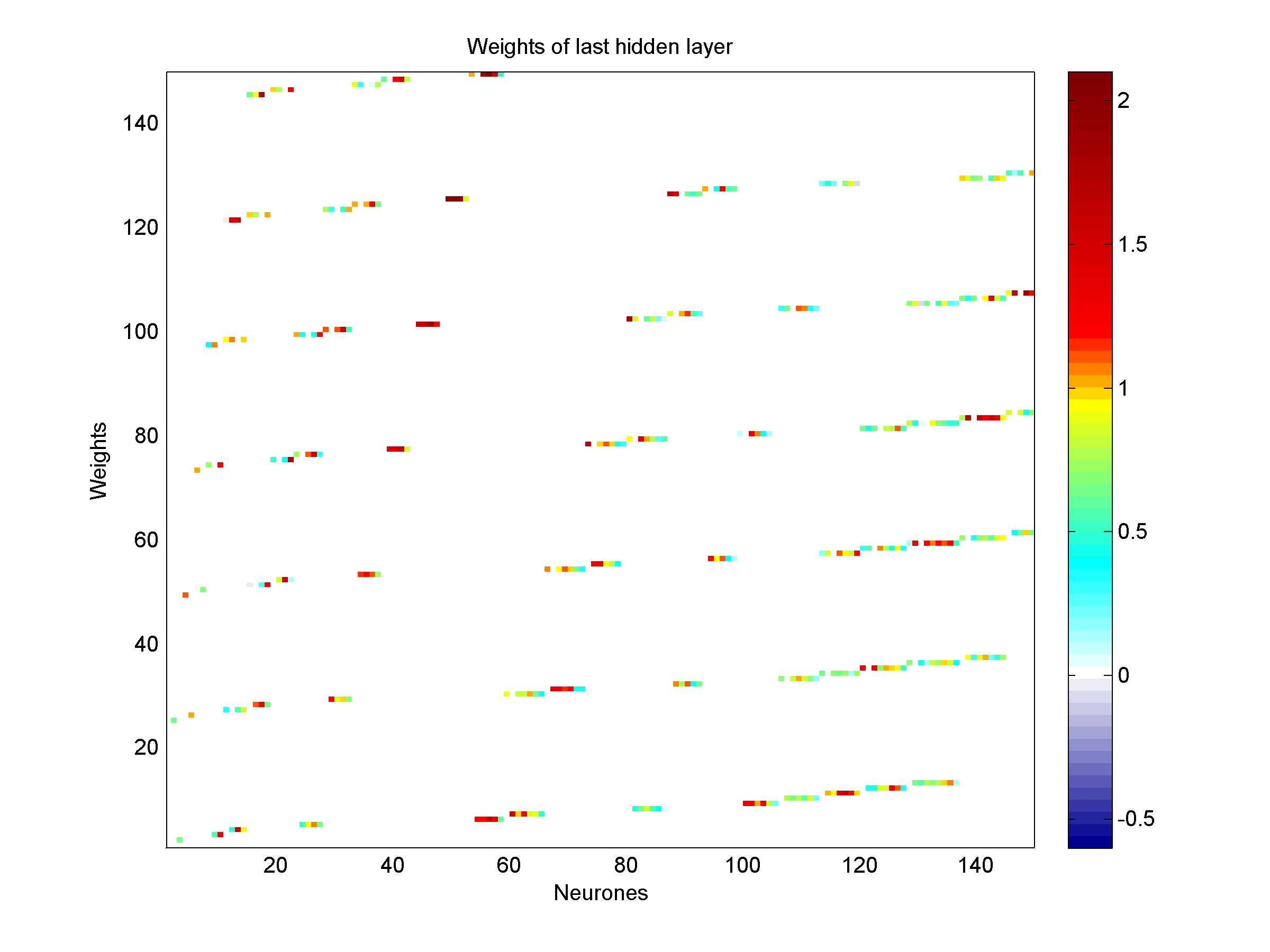}
	\caption{Weights of the last hidden layer in deep neural network for BCH(63,45) code.}
	\label{fig:layer4_dl}
\end{figure}

We compared the weights of the BP algorithm and the weights of the trained deep neural network for a BCH(63,45) code. We observed that the deep neural network produces weights in the range from $-0.8$ to $2.2$, in contrast to the BP algorithm which has binary $“1”$ or $“0”$ weights. Figure~\ref{fig:weight_hist} shows the weights histogram for the last layer. Interestingly, the distribution of the weights is close to a normal distribution. In a similar way, every hidden layer in the trained deep neural network has a close to normal distribution. Note that Hinton \cite{hinton2010practical} recommends to initialize the weights with normal distribution.
In Figures~\ref{fig:layer4_bp} and~\ref{fig:layer4_dl} we plot the weights of the last hidden layer. Each column in the figure corresponds to a neuron described by Equation ~\eqref{eq:x_ie_RB_NN}. It can be observed that most of the weights are zeros except the Tanner graph weights which have value of 1 in Figure~\ref{fig:layer4_bp} (BP algorithm) and some real number in Figure~\ref{fig:layer4_dl} for the neural network. In Figure~\ref{fig:layer4_bp} and ~\ref{fig:layer4_dl} we plot a quarter of the weights matrix for better illustration.

\section{CONCLUSIONS}
In this work we applied deep learning techniques to improve the performance of the BP algorithm. We showed that a ``soft'' Tanner graph can produce improvements when used in the BP algorithm instead of the standard Tanner graph. We believe that the BER improvement was achieved by properly weighting the messages, such that the effect of small cycles in the Tanner graph was partially compensated. It should be emphasized that the parity check matrices that we worked with were obtained from \cite{ParityCheckMatrix}. We have not evaluated our method on parity check matrices for which an attempt was made to reduce the number of small cycles. A notable property of our neural network decoder is that once we have trained its parameters, we can improve performance compared to plain BP without increasing the required computational complexity. Another notable property of the neural network decoder is that we learn the channel and the linear code simultaneously. 

We regard this work as a first step in the implementation of deep learning techniques for the design of improved decoders. Our future work include possible improvements in the error rate results by exploring new neural network architectures and combining other decoding methods. Furthermore, we plan to investigate the connection between the parity check matrix and the deep neural network decoding capabilities.

\addtolength{\textheight}{-12cm}   





\section*{ACKNOWLEDGMENT}

This research was supported by the Israel Science Foundation, grant no. 1082/13. The Tesla K40c used for this research was donated by the NVIDIA Corporation.


\bibliographystyle{IEEEtran}
\bibliography{IEEEexample}




\end{document}